\documentclass[12pt]{article}
\usepackage{epsf}
\begin{document}

\begin{titlepage}

\rightline{hep-ph/0207189}
\rightline{July, 2002}

\begin{center}

{\Large\bf  Vector-meson contributions do not explain the rate and
spectrum in \mbox{\boldmath $K_L \rightarrow \pi^0 \gamma \gamma$} }

\medskip

\normalsize
{\large F. Gabbiani$^{}$ and G. Valencia$^{}$} \\
\vskip .3cm
$^{}$Department of Physics and Astronomy, Iowa State University,
Ames, Iowa 50011\\
\vskip .3cm

\end{center}

\begin{abstract}

We analyze the recent NA48 data for the reaction
$K_L \rightarrow \pi^0 \gamma \gamma$ with and without the
assumption of vector meson dominance (VMD). We find that
the data are well described by a three-parameter expression
inspired by ${\cal O}(p^6)$ chiral perturbation theory.
We also find that it is impossible to fit the shape of the
decay distribution and the overall rate simultaneously if one
imposes the VMD constraints on the three parameters.
We comment on the different fits and their implications for the
$CP$-conserving component of the decay $K_L \rightarrow \pi^0 e^+ e^-$.

\end{abstract}

\end{titlepage}

\section{Introduction}

In Ref.~\cite{gabval} we examined the KTeV \cite{ktev} data for
the mode $K_L \rightarrow \pi^0 \gamma \gamma$ using a more general
(three-parameter) description than the one used by KTeV. The latter
has become the norm in the literature and it follows from an assumption
of vector meson dominance (VMD) \cite{vmdref} in conjunction with
the parametrization inspired by ${\cal}O(p^6)$ chiral perturbation
theory of Ref.~\cite{unitco}. We argued that VMD
in this decay is an experimental question and, therefore, that it
should not be an input to the data analysis. We found a least squares
best fit to the data within our approach
that was slightly better than the usual fit. However, it was hard to
reach definitive conclusions because the necessary information is not made
available by KTeV. Nevertheless, we motivated our more general approach
by showing that there are important contributions to this decay from
intermediate $f^{\phantom{l}}_2(1270)$ mesons
that do not conform to the VMD parametrization \cite{other}.

In this paper we present our three-parameter fit for newly released
data from NA48 \cite{na48}. This is important for the following reasons.
First, the NA48 data are significantly
different from the KTeV data and leads to different
conclusions regarding the $CP$-conserving contribution to
$K_L \rightarrow \pi^0 e^+ e^-$~\cite{kpeeold}--\cite{doga}.
Second, NA48 has presented their data in
a form that allows us to directly compare our general fit to the
usual VMD fit. This allows us to show that whereas it is possible
to fit the decay distribution $d\Gamma/dm_{\gamma\gamma}$ equally well
with the general and VMD approaches, only the former is capable of
fitting simultaneously the decay distribution and the total decay rate.

We also discuss two additional issues. First, we show that the
two fits to the decay distribution (the general and the VMD
parametrizations) correspond respectively to constructive and
destructive interference between two amplitudes.
Second, we comment on the dependence of the
fit on the parameter $a^{\phantom{l}}_2$, which is extracted from
$K \rightarrow \pi\pi\pi$ decays and which has a large uncertainty
\cite{k3pi,bijnens}.

\section{Parametrization of the data}

The $K_L\rightarrow\pi^0\gamma\gamma$ amplitude in the limit of
$CP$ violation can be written in terms of two independent invariant
amplitudes, A and B \cite{rafael},
\begin{eqnarray}
\lefteqn{{\cal M}[K_L(p^{\phantom{l}}_K) \rightarrow \pi^0(p^{\phantom{l}}_\pi)
\gamma(q^{\phantom{l}}_1)\gamma(q^{\phantom{l}}_2)]\, =\,
{G^{\phantom{l}}_8 \alpha^{\phantom{l}}_{EM} \over 4
 \pi}\epsilon^{\phantom{l}}_\mu(q^{\phantom{l}}_1)
\,\epsilon^{\phantom{l}}_\nu(q^{\phantom{l}}_2) \, \Bigg [
    A\,
    \left(q_2^\mu q_1^\nu - q^{\phantom{l}}_1\cdot
q^{\phantom{l}}_2 \, g^{\mu\nu}\right) } \nonumber \\
&&
+ 2 {B \over {m^2_K}}\,
\left(p^{\phantom{l}}_K\cdot q^{\phantom{l}}_1 \, q_2^\mu p_K^\nu
+ p^{\phantom{l}}_K\cdot q^{\phantom{l}}_2\, q_1^\nu p_K^\mu
- q^{\phantom{l}}_1\cdot q^{\phantom{l}}_2 \, p_K^\mu p_K^\nu -
p^{\phantom{l}}_K\cdot q^{\phantom{l}}_1\, p^{\phantom{l}}_K\cdot
q^{\phantom{l}}_2 \, g^{\mu\nu}\right) \Bigg ]\;.
\label{gen}
\end{eqnarray}
The Fermi constant and the Cabibbo angle are included in the overall
constant $G^{\phantom{l}}_8 = 9.1 \times 10^{-6}$~GeV$^{-2}$ and
$\alpha^{\phantom{l}}_{EM} \approx 1/137$ is the usual electromagnetic
fine structure constant. To parametrize these amplitudes in a
form inspired by ${\cal O}(p^6)$ chiral perturbation theory and
dispersion relations Ref.~\cite{unitco} proposed the use of:
\begin{eqnarray}
A(z,y) & = & 4 F\left(\frac{z}{r^2_{\pi}}\right) {a^{\phantom{l}}_1(z) \over z} +4 {F(z) \over z}
(1+r^2_{\pi}-z) \nonumber \\
& + &  {a^{\phantom{l}}_2 M^2_K \over {\Lambda^2_{\chi}}} \left\{ {4
r^2_{\pi} \over z} F\left(\frac{z}{r^2_{\pi}}\right) + {2 \over 3} \left(2 +
\frac{z}{r^2_{\pi}}\right) \left [{1 \over 6} + R\left(\frac{z}{r^2_{\pi}}\right)
\right] - {2 \over 3} \log {m^2_{\pi} \over M^2_{\rho}} \right.
\nonumber  \\
&-& 2 \frac{r^2_{\pi}}{z^2}(z+1-r^2_{\pi})^2 \left[ \frac{z}{12 r^2_{\pi}} +
F\left(\frac{z}{r^2_{\pi}}\right) + \frac{z}{r^2_{\pi}}
R\left(\frac{z}{r^2_{\pi}}\right)\right] \nonumber \\
&+& \left. 8 \frac{r^2_{\pi}}{z^2} y^2
\left[\frac{z}{12 r^2_{\pi}} + F\left(\frac{z}{r^2_{\pi}}\right) + \frac{z}{2
r^2_{\pi}} F\left(\frac{z}{r^2_{\pi}}\right) + 3 \frac{z}{r^2_{\pi}}
R\left(\frac{z}{r^2_{\pi}}\right) \right ] \right \} \nonumber  \\
& + &
\alpha^{\phantom{l}}_1 (z-r^2_{\pi})+\alpha^{\phantom{l}}_2\;, \nonumber \\
B(z) & = & {a^{\phantom{l}}_2 M^2_K \over {\Lambda^2_{\chi}}} \left\{ {4 r^2_{\pi} \over z}
F\left(\frac{z}{r^2_{\pi}}\right) + {2 \over 3} \left(10 - \frac{z}{r^2_{\pi}}\right)
\left[ {1 \over 6} + R\left (\frac{z}{r^2_{\pi}}\right) \right] + {2 \over 3}
\log {m^2_{\pi} \over m^2_{\rho}} \right\} +
\beta\;,
\label{ampl}
\end{eqnarray}
where the dimensionless kinematic variables are
\begin{equation}
z  =  { \left( q^{\phantom{l}}_1 + q^{\phantom{l}}_2 \right)^2 \over M^2_K}\; ,\ \
y = {p^{\phantom{l}}_K \cdot(q^{\phantom{l}}_1-q^{\phantom{l}}_2)\over M^2_K}\;,
\end{equation}
and the scale of chiral symmetry breaking is
$\Lambda^{\phantom{l}}_{\chi}$ $\approx$ $4\pi
f^{\phantom{l}}_{\pi}$ $\approx$ 1.17 GeV.

The form in Eq.~(\ref{ampl}) does not correspond to a complete calculation
at order $p^6$ in chiral perturbation theory. Rather it contains the complete
one-loop calculation at order $p^4$ \cite{locpt} and two additional
ingredients containing some corrections of order $p^6$ \cite{unitam,unitco}.
The non-analytic terms
in Eq.~(\ref{ampl}) that multiply the factors $a^{\phantom{l}}_2$
and $a^{\phantom{l}}_1(z)$ attempt to incorporate the strong
rescattering in the two-pion intermediate state that occurs at one-loop.
They arise from the inclusion of $p^4$ corrections to the
$K \rightarrow 3 \pi$ amplitudes \cite{k3pi,dghkp}. The values of
$a^{\phantom{l}}_1$ and $a^{\phantom{l}}_2$ are extracted from data and
the functions $F(z)$ and $R(z)$ can be
found in the literature \cite{unitco}.
The three constants $\alpha^{\phantom{l}}_1$,
$\alpha^{\phantom{l}}_2$ and $\beta$ originate in
counterterms appearing in the $p^6$ weak chiral Lagrangian \cite{unitco}.

The analysis of $K \rightarrow 3 \pi$ in
Ref. ~\cite{k3pi} indicates that
\begin{eqnarray}
a^{\phantom{l}}_1(z) =  0.38 + 0.13 \; Y^{\phantom{l}}_0 - 0.0059 \; Y^2_0\;, &&
a^{\phantom{l}}_2  =  6.5\;, \nonumber \\
Y^{\phantom{l}}_0 =  {(z-r^2_{\pi} - {1 \over 3})\over r^2_{\pi}}\;, &&
r^{\phantom{l}}_\pi=m^{\phantom{l}}_\pi/M^{\phantom{l}}_K\;.
\end{eqnarray}
A very recent analysis of $K\rightarrow 3 \pi$ data results in \cite{bijnens}
\begin{equation}
a^{\phantom{l}}_2 =  6.8 \pm 2.4\;.
\label{newa2}
\end{equation}

In the analysis of Ref. \cite{unitco}, which has become standard,
the three unknown constants were fixed in
terms of the contribution they receive from vector-meson exchange,
supplemented with a minimal subtraction ansatz:
\begin{eqnarray}
\alpha^{\phantom{l}}_1 &=& -4 a^{\phantom{l}}_V\;, \nonumber \\
\alpha^{\phantom{l}}_2 &=& 12 a^{\phantom{l}}_V -0.65\;, \nonumber \\
\beta &=& -8 a^{\phantom{l}}_V - 0.13\;,
\label{cohenan}
\end{eqnarray}
and this form has been used both by KTeV \cite{ktev} and by NA48 \cite{na48}
to fit their data. In Ref.~\cite{gabval} we argued that this ansatz
imposes a correlation on $\beta$ that is not desirable for a
prediction of the $CP$-conserving contribution to
$K_L \rightarrow \pi^0 e^+ e^-$. With
the new NA48 data we can go further and conclude that the VMD ansatz
does not provide a good description of $K_L \rightarrow \pi^0 \gamma \gamma$.

\section{Fitting the shape of the \mbox{\boldmath $\lowercase{d}\Gamma/\lowercase{dm_{\gamma\gamma}}$
distribution}}

NA48 has recently released their result for
$K_L \rightarrow \pi^0 \gamma \gamma$ \cite{na48}. They chose to
analyze their data using Eq.~(\ref{ampl}) with
the VMD  assumption, and they found
$a^{\phantom{l}}_V = -0.46$. To obtain this number they
fit the shape of the distribution $d\Gamma /dy dm_{\gamma\gamma}$
without attempting to fit the branching ratio. NA48 has published
in Table~2 of Ref. \cite{na48} sufficient information to fit the
distribution $d\Gamma /dm_{\gamma\gamma}$. They present
the number of unambiguous events, estimated background and acceptance for
each 20 MeV bin in $m^{\phantom{l}}_{\gamma\gamma}$.

We begin our analysis with a fit to the shape of the $d\Gamma /dm_{\gamma\gamma}$
distribution, ignoring the measured branching ratio, to compare with the
fit performed by NA48. We do this both using the VMD assumption and with the
general approach. We calculate the number of events predicted in each bin
as
\begin{equation}
N_i = N \biggl[{1 \over \Gamma_{K_L}} \int_i dm_{\gamma\gamma}
\biggl({d\Gamma \over dm_{\gamma\gamma}}\biggr) N(K_L) \biggr] {\rm Acceptance}_i
+ {\rm Background}_i \;,
\label{fitshape}
\end{equation}
where $N$ is a normalization chosen to match the total number of
events and $N(K_L)$ = 23.9 $\times$ 10$^9$ is the number of decays in
the fiducial volume. The arbitrary normalization allows us to fit the shape of
the distribution while ignoring the overall rate.

We use data from 17 out of 20 bins presented in Table~2
of Ref.~\cite{na48}. We exclude two bins in
the $m^{\phantom{l}}_{\gamma\gamma}$ region near the $\pi^0$ mass which do not have any
events due to kinematic cuts, and we also exclude the last bin with no
events because it lies outside the physical region. We perform a least
squares fit using Poisson statistics for the bins with small number of
events following  Ref.~\cite{stat}.

\begin{figure}[!htb]
\begin{center}
\epsfxsize=15cm
\centerline{\epsffile{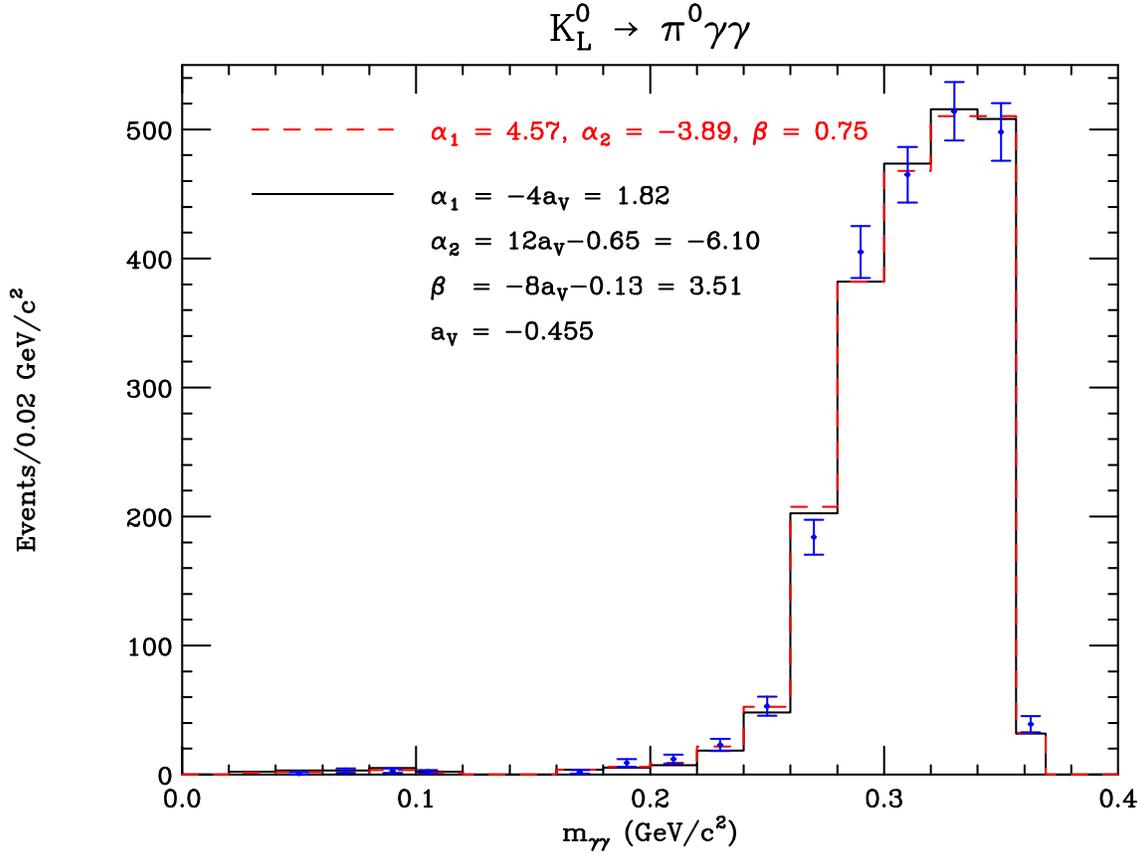}}
\end{center}
\caption{Two different fits to the data from Ref.~\cite{na48}, as
explained in the text. The solid line is a one-parameter fit
corresponding to Eq.~(\ref{avfit}), the dashed line is the three-parameter
fit shown in Eq.~(\ref{3pfit}).}
\label{fig:fitna48a}
\end{figure}

With this procedure, and the VMD ansatz, we reproduce approximately the
NA48 best fit. We obtain $a^{\phantom{l}}_V = -0.455$
with a $\chi^2/dof = 18.5/16$. We show this result in
Fig.~\ref{fig:fitna48a} where we superimpose our best three-parameter
fit which has a $\chi^2/dof = 14.6/14$.  The two fits are nearly
identical as can be seen in the figure and they are indistinguishable
statistically. Nevertheless, when they are both expressed in terms of
the three general parameters one can see they correspond to very
different solutions. For the general fit,
\begin{eqnarray}
\alpha^{\phantom{l}}_1 = 4.57\;,
\qquad \alpha^{\phantom{l}}_2 = -3.89\;,
\qquad \beta = 0.75\; ;
\label{3pfit}
\end{eqnarray}
whereas for the VMD fit (in terms of $a^{\phantom{l}}_V$),
\begin{eqnarray}
\alpha^{\phantom{l}}_1 = 1.82\;, \qquad
\alpha^{\phantom{l}}_2 = -6.10\;, \qquad
\beta = 3.51\;.
\label{avfit}
\end{eqnarray}

For the case of the three-parameter fit we find
that $\alpha^{\phantom{l}}_1$ and $\alpha^{\phantom{l}}_2$ are correlated as
was discussed in Ref.~\cite{gabval}, so that there are many other fits
with a $\chi^2$ near the minimum for the same value of $\beta$.

As stated above, neither one of these fits reproduces the experimental
rate, $B(K_L \rightarrow \pi^0 \gamma \gamma) = (1.36 \pm 0.03 \pm 0.03)
\times 10^{-6}$ \cite{na48}. The theoretical branching ratio predicted
for $a^{\phantom{l}}_V = -0.455$ (the NA48 value) is
$B(K_L \rightarrow \pi^0 \gamma \gamma) = 1.1 \times 10^{-6}$, and the
one predicted for the three parameters in  Eq.~(\ref{3pfit}) is
$B(K_L \rightarrow \pi^0 \gamma \gamma) = 1.0 \times 10^{-6}$.

\begin{figure}[!htb]
\begin{center}
\epsfxsize=15cm
\centerline{\epsffile{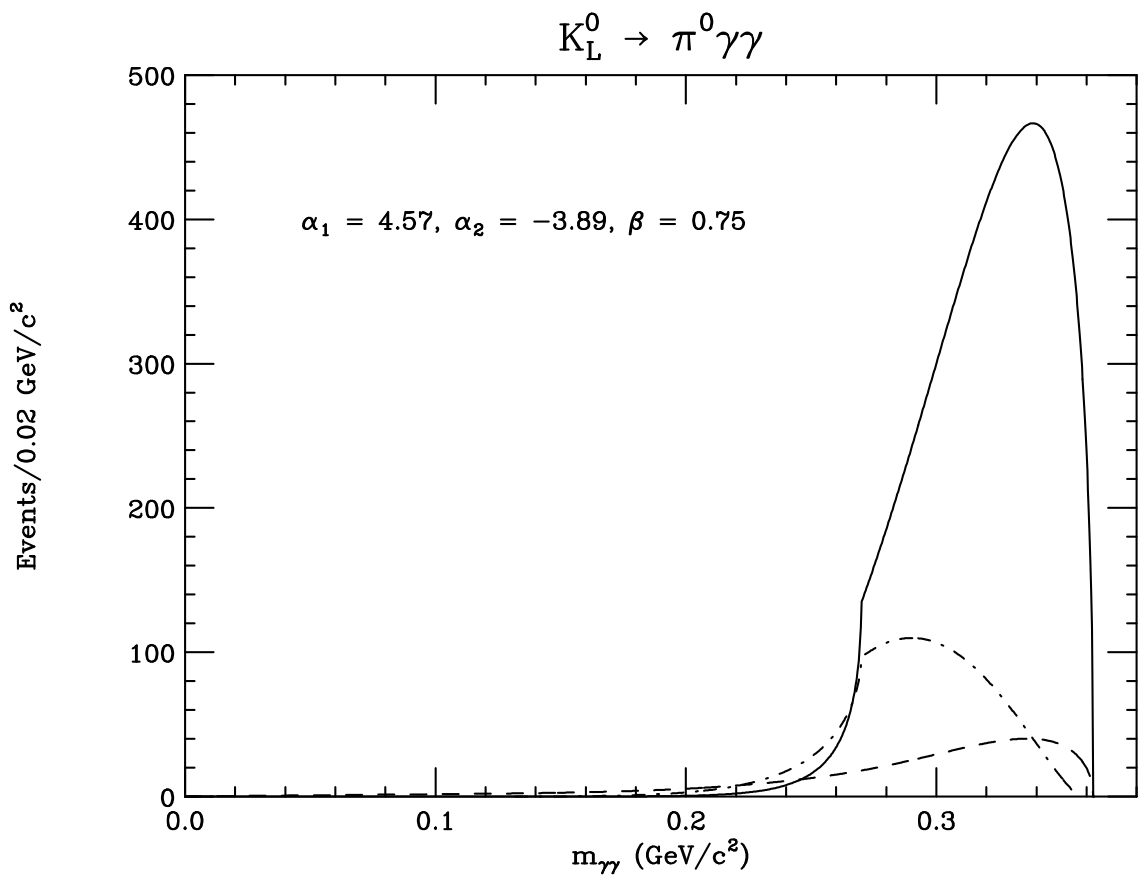}}
\end{center}
\caption{The contributions from $\vert A \vert^2$ (solid line), $\vert B
\vert^2$ (dashed line), and Re($A^{\star} B$) (dot-dashed line) are
plotted vs. the invariant two-photon mass $m_{\gamma \gamma}$ in terms
of the number of events for the best three-parameter fit.}
\label{fig:separ1}
\end{figure}

\begin{figure}[!htb]
\begin{center}
\epsfxsize=15cm
\centerline{\epsffile{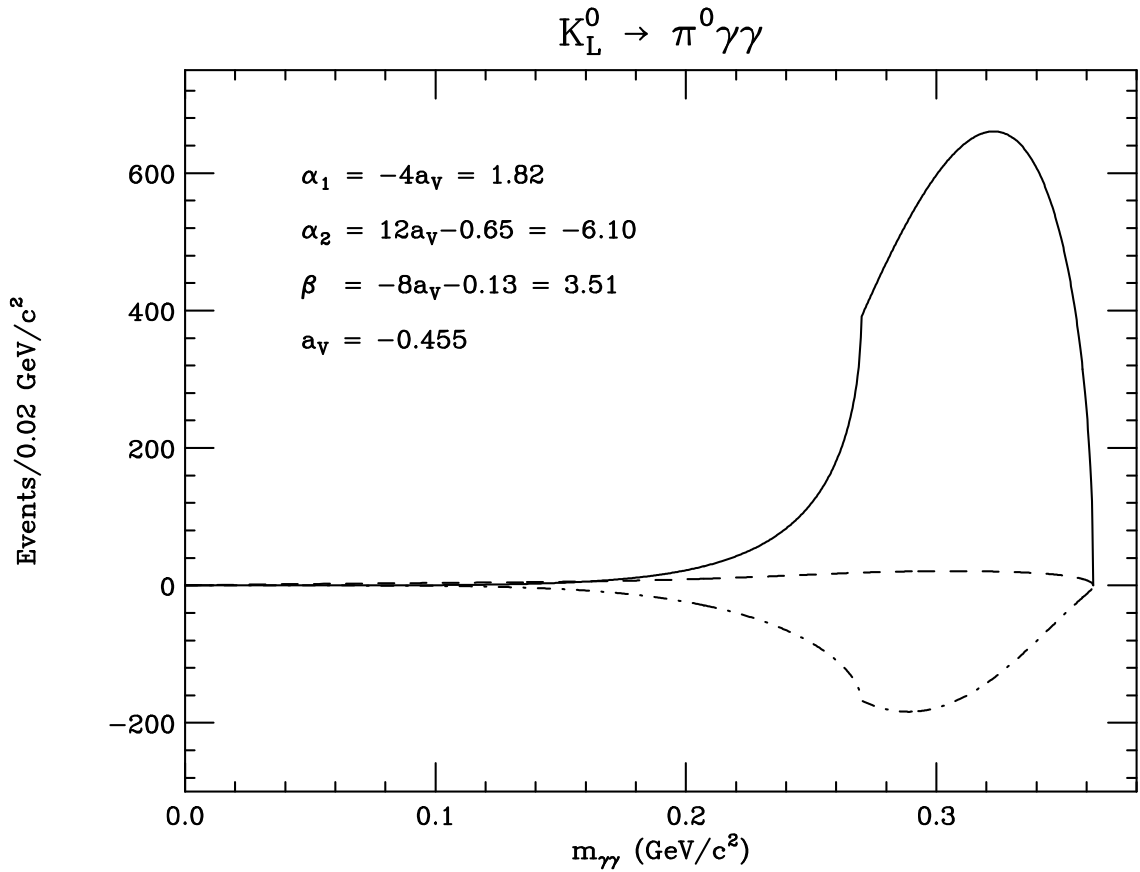}}
\end{center}
\caption{Same as in Fig.~\ref{fig:separ1} for the best $a_V$ fit.}
\label{fig:separ2}
\end{figure}

It is instructive to show the three separate contributions that result
from Eq.~(\ref{ampl}) to the differential decay rate
$dB(K_L \rightarrow \pi^0 \gamma \gamma)/dm_{\gamma\gamma}$. The three
terms correspond to the absolute square of the $A$ and $B$ amplitudes
and to their interference, $|A|^2$, $|B|^2$ and ${\rm Re}(A^\star B)$,
respectively.
We show these quantities in
Fig.~\ref{fig:separ1} for the best three-parameter fit and
Fig.~\ref{fig:separ2} for the best $a^{\phantom{l}}_V$ fit.

In both of these figures the solid line represents the contribution
from $|A|^2$, the dashed line the contribution from $|B|^2$ and the
dot-dashed line the interference. We observe that the three-parameter
fit corresponds to constructive interference between the $A$ and $B$
amplitudes, whereas the $a^{\phantom{l}}_V$ fit corresponds to destructive
interference. Unfortunately it appears that it is not possible to
determine experimentally the sign of this interference. However, as
we show below, the total rate for the process discriminates between
the VMD ansatz and the general form of the amplitude.

\section{Simultaneous fit to the shape of the \mbox{\boldmath
$d\Gamma/dm_{\gamma\gamma}$} distribution and to the decay rate}

\begin{figure}[!htb]
\begin{center}
\epsfxsize=15cm
\centerline{\epsffile{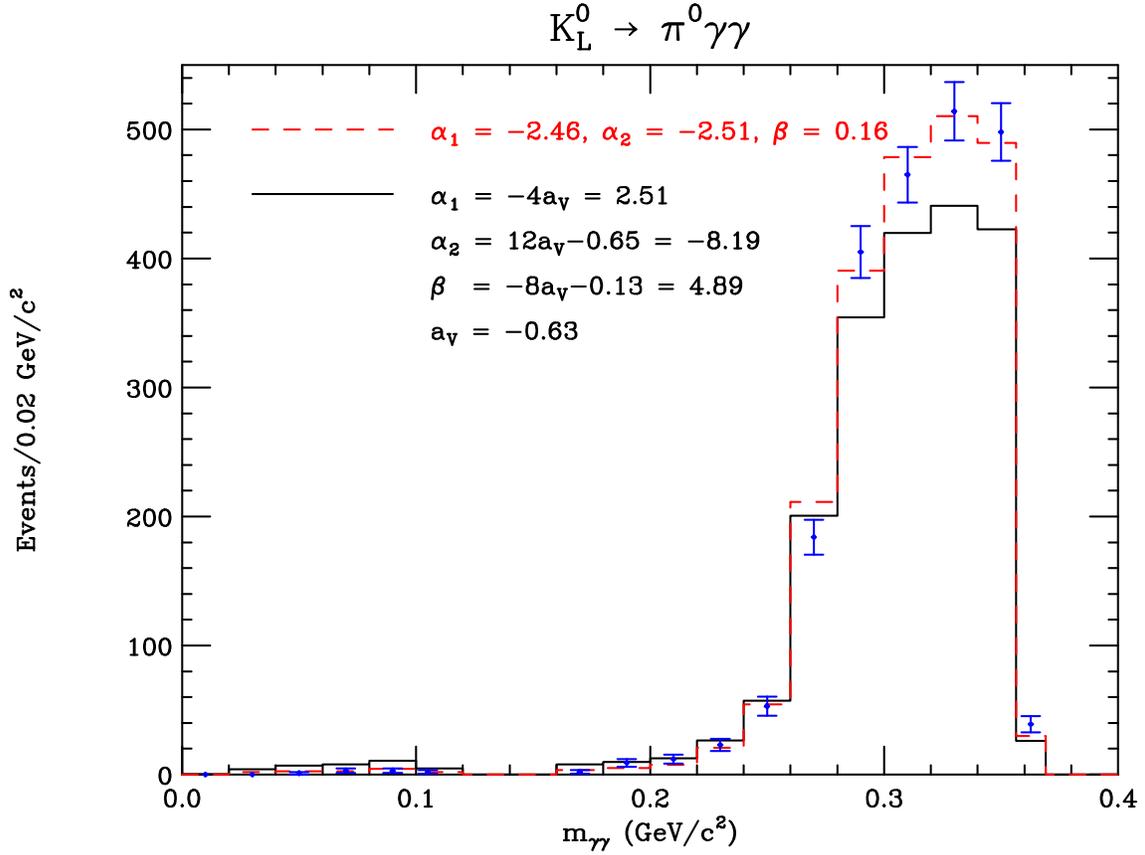}}
\end{center}
\caption{A simultaneous fit to
the shape of $d\Gamma/dm_{\gamma\gamma}$ and to the decay
rate. The solid line is a one-parameter fit corresponding to
Eq.~(\ref{avfitr}), the dashed line is the three-parameter fit shown
in Eq.~(\ref{3pfitr}).}
\label{fig:fitna48b}
\end{figure}

\begin{figure}[!htb]
\begin{center}
\epsfxsize=15cm
\centerline{\epsffile{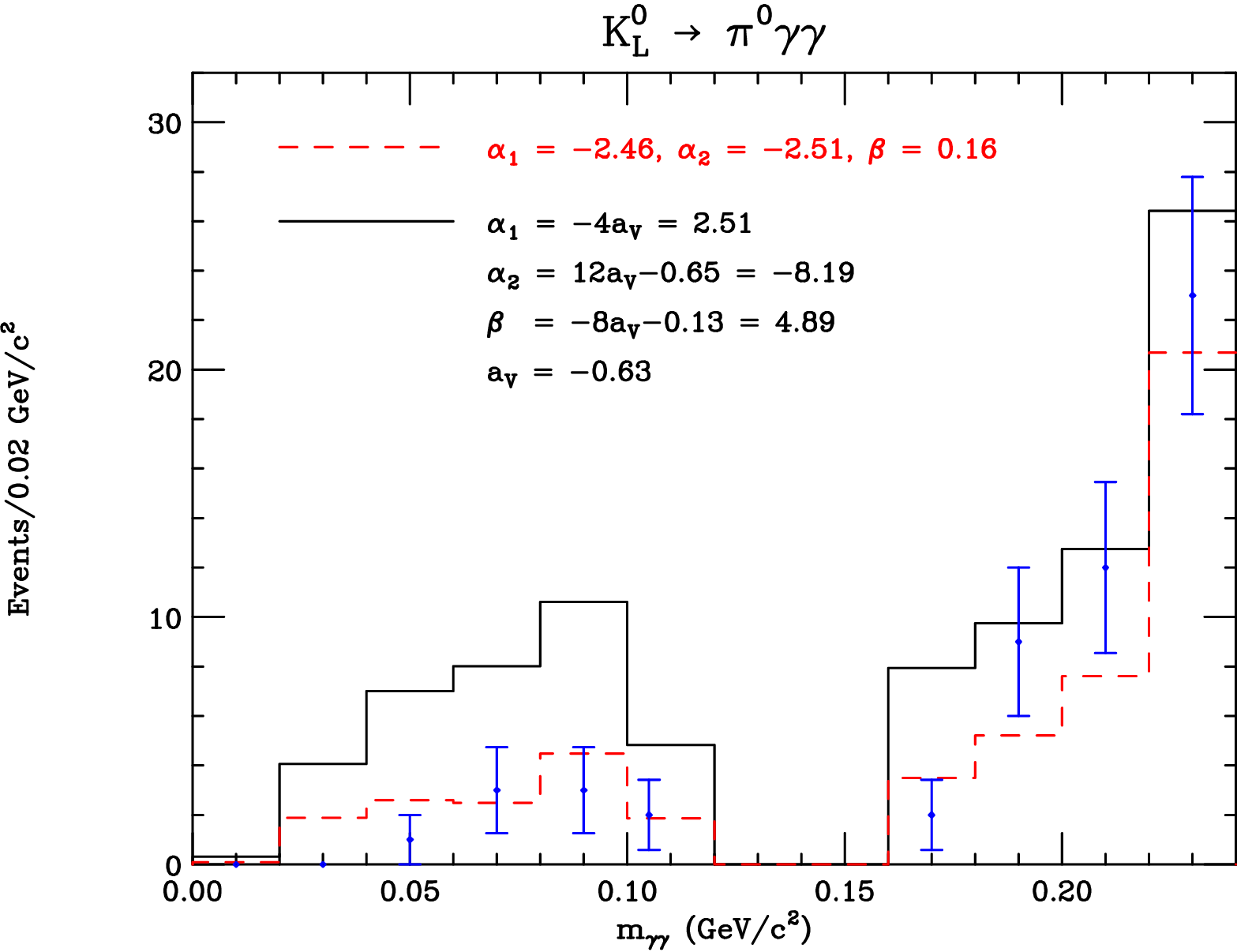}}
\end{center}
\caption{An enlargement of Fig.~\ref{fig:fitna48b} for
$m_{\gamma\gamma} \leq 0.24$ GeV/c$^2$.} 
\label{fig:fitna48c}
\end{figure}

To obtain a fit that reproduces the observed branching ratio we proceed
as in Eq.~(\ref{fitshape}) but removing the arbitrary normalization,
\begin{equation}
N_i =  \biggl[{1 \over \Gamma_{K_L}} \int_i dm_{\gamma\gamma}
\biggl({d\Gamma \over dm_{\gamma\gamma}}\biggr) N(K_L) \biggr] {\rm Acceptance}_i
+ {\rm Background}_i \;,
\label{fitrate}
\end{equation}
with the same notation of Eq.~(\ref{fitshape}).
We first attempt this fit with the VMD ansatz and find that it is impossible
to obtain a good fit. Our least squares fit using the VMD ansatz
occurs for $a^{\phantom{l}}_V = -0.63$ and has a $\chi^2/dof = 74.8/16$.
We show this result as the solid line in Figs.~\ref{fig:fitna48b} and \ref{fig:fitna48c}.
The implied branching ratio is
$B(K_L \rightarrow \pi^0 \gamma \gamma) = 1.25 \times 10^{-6}$ and
$a^{\phantom{l}}_V = -0.63$ corresponds to
\begin{eqnarray}
\alpha^{\phantom{l}}_1 = 2.51\;, \qquad
\alpha^{\phantom{l}}_2 = -8.19\;, \qquad
\beta = 4.89\;.
\label{avfitr}
\end{eqnarray}

Our best three-parameter fit, on the other hand, has a
$\chi^2/dof = 18.5/14$ and is shown as the dashed line in
Figs.~\ref{fig:fitna48b} and \ref{fig:fitna48c}.
It implies a branching ratio
$B(K_L \rightarrow \pi^0 \gamma \gamma) = 1.36 \times 10^{-6}$ in good
agreement with the measured one. The parameters for this best fit are,
\begin{eqnarray}
\alpha^{\phantom{l}}_1 = -2.46\;,
\qquad \alpha^{\phantom{l}}_2 = -2.51\;,
\qquad \beta = 0.16\; .
\label{3pfitr}
\end{eqnarray}
We conclude from Fig.~\ref{fig:fitna48b} that the VMD ansatz cannot reproduce
the shape of the spectrum and the total decay rate simultaneously, but that the
general formula, Eq.~(\ref{ampl}) does accommodate both.

\begin{figure}[!htb]
\begin{center}
\epsfxsize=15cm
\centerline{\epsffile{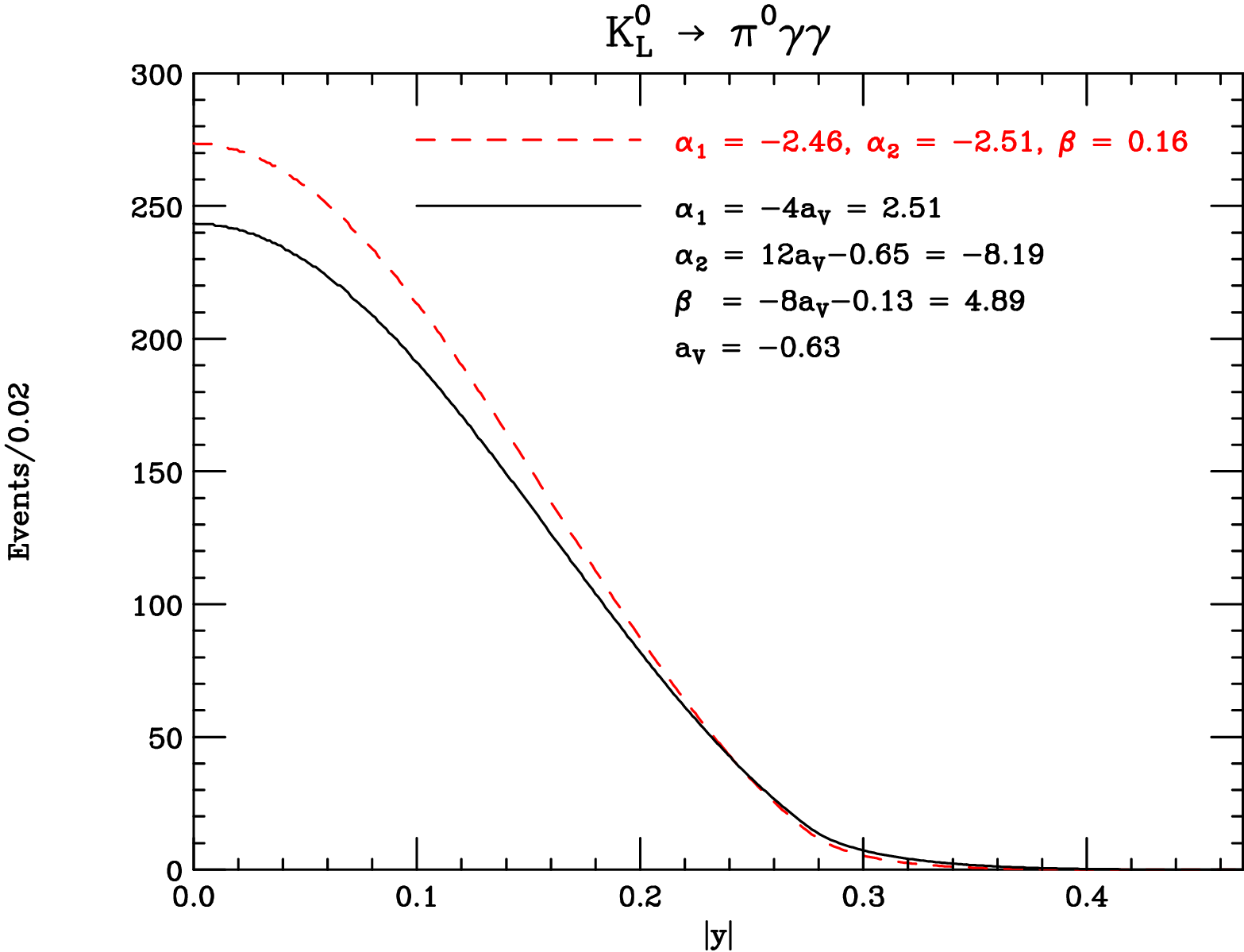}}
\end{center}
\caption{Theoretical $d\Gamma/dy$ distributions.
The solid and dashed lines are predicted using as input the results given in
Eq.~(\ref{avfitr}) and Eq.~(\ref{3pfitr}), respectively.}
\label{fig:y}
\end{figure}

\begin{figure}[!htb]
\begin{center}
\epsfxsize=15cm
\centerline{\epsffile{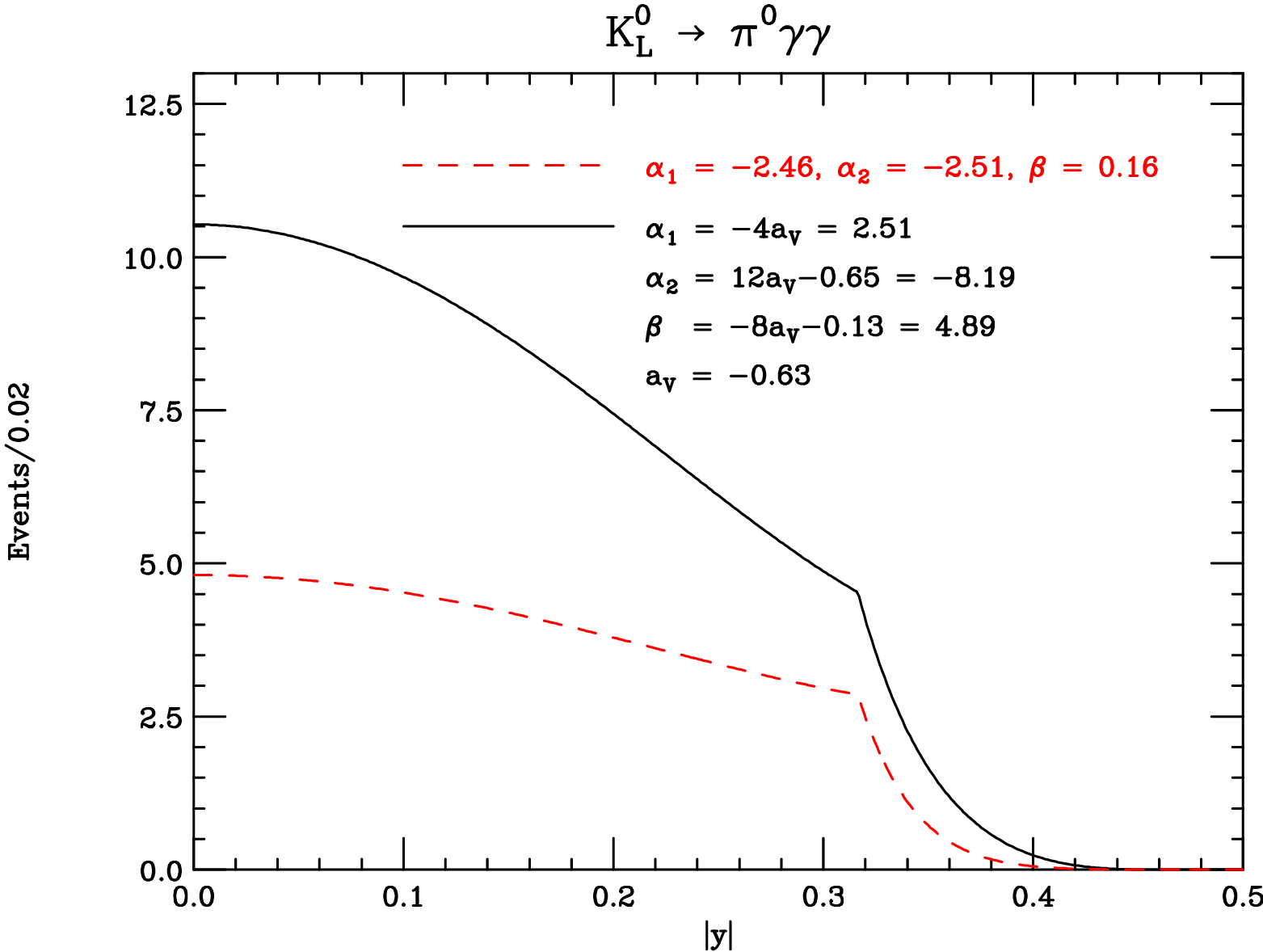}}
\end{center}
\caption{Same as Fig.~\ref{fig:y} but with a cut at $m_{\gamma\gamma} \leq
  0.24$ GeV/c$^2$.}
\label{fig:yc}
\end{figure}

For completeness we show in Figs.~{\ref{fig:y}} and {\ref{fig:yc}} the theoretical
$d\Gamma/dy$ distributions for both the $a^{\phantom{l}}_V$ result
from Eq.~(\ref{avfitr}) and the three parameters given in
Eq.~(\ref{3pfitr}). Fig.~{\ref{fig:yc}} is restricted to events with
$m_{\gamma\gamma} \leq 0.24$ GeV/c$^2$.
There are no data available in this form, so at
this point we are not able to perform a fit and we can only present
our predictions. We point out that the three-parameter fit yields a
flatter distribution than the $a^{\phantom{l}}_V$ fit.

\subsection{Dependence on \mbox{\boldmath $a_2$}}

We now consider the dependence of our results on the parameter
$a^{\phantom{l}}_2$ that appears in the $B$ amplitude.
This parameter is extracted
from $K \rightarrow 3 \pi$ decays and up to now we have used
the value $a^{\phantom{l}}_2 = 6.5~\cite{unitco}$.
However, the value of this parameter has a large uncertainty, of order
$\sim 35 \, \%$. For example, from the recent analysis of Ref.~\cite{bijnens}
one extracts $a^{\phantom{l}}_2 = 6.8 \pm 2.4$.

The analytic form for the $B$ amplitude in Eq.~(\ref{ampl})
clearly indicates that $a^{\phantom{l}}_2$ and $\beta$ are
correlated and this is confirmed by our
numerical study. It is possible to obtain many
equally good fits to the data with different values of $a^{\phantom{l}}_2$ and
$\beta$. For example if we take the central value from Ref.~\cite{bijnens}
and 1-sigma deviations from it, we find good fits to the shape and spectrum
with the values listed in Table~\ref{tab}. This is not possible with the
$a^{\phantom{l}}_V$ parametrization, where we cannot find a good fit for any of these
values of $a^{\phantom{l}}_2$.

\begin{table}[htb]
\centering
\begin{tabular}{|c|c|c|c|c|} \hline
$a^{\phantom{l}}_2$ & $\alpha^{\phantom{l}}_1$ &
  $\alpha^{\phantom{l}}_2$ & $\beta$ & $\chi^{2^{\phantom{l}}}/dof$ \\
\hline
\hline
6.8 & --2.42 & --2.65 & 0.25 & 18.5/14 \\
\hline
4.4 & --2.33 & --1.71 & --0.46 & 18.4/14 \\
\hline
9.2 & --2.58 & --3.51 & 0.91 & 18.6/14 \\
\hline
\end{tabular}
\vskip 0.1in
\caption{Three-parameter best fits for three different values of
$a_2$, corresponding to its central value from
Ref.~\cite{bijnens} and its 1-sigma deviations.}
\label{tab}
\end{table}

\section{\mbox{\boldmath $CP$}-conserving contribution to
\mbox{\boldmath $K_L \rightarrow \pi^0 e^+ e^-$}}

\begin{figure}[!htb]
\begin{center}
\epsfxsize=15cm
\centerline{\epsffile{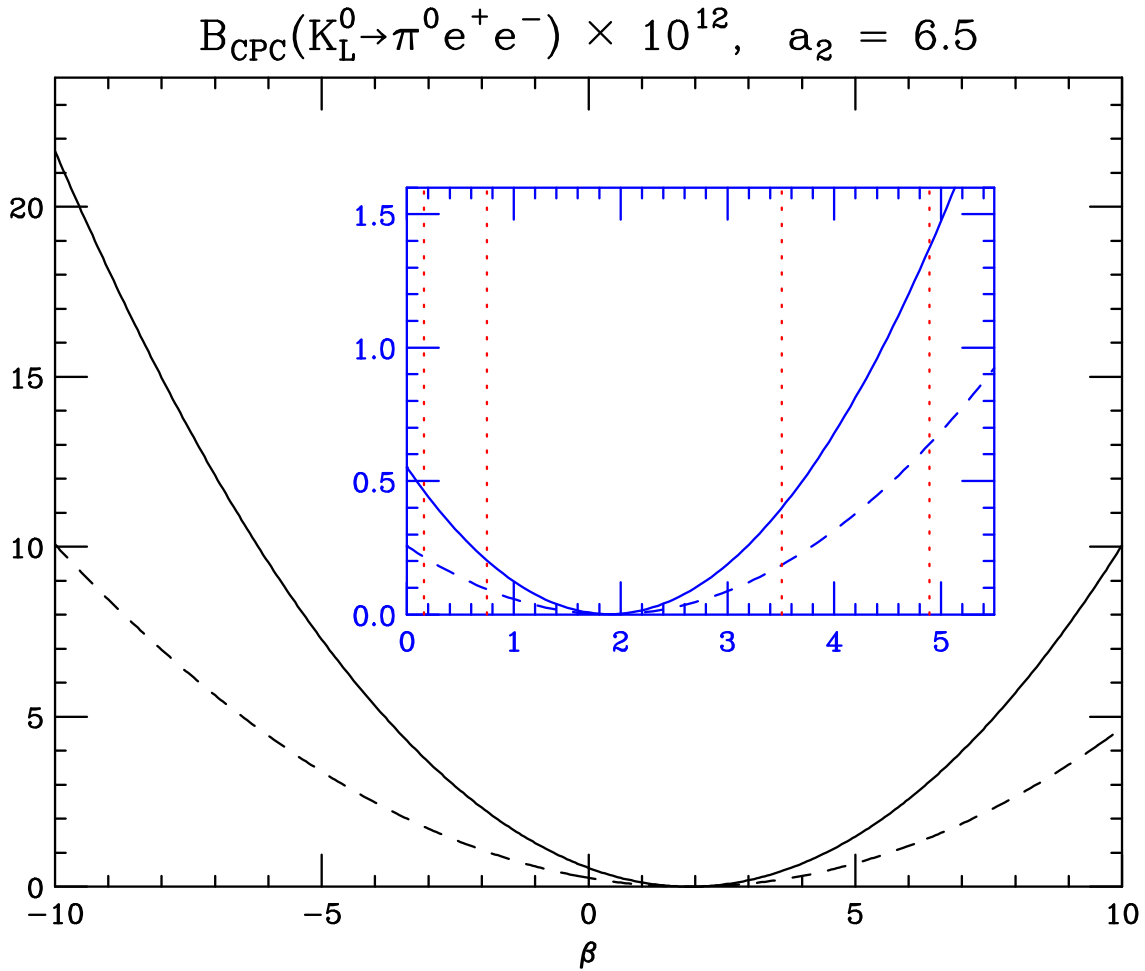}}
\end{center}
\caption{$CP$-conserving contribution to $K_L \rightarrow \pi^0 e^+ e^-$
as a function of $\beta$ with $a_2$ = 6.5~\cite{unitco}. The dashed line shows the
absorptive contribution and the solid line the model of
Ref.~\cite{doga}. The enlargement shows the results for
the branching ratio vs. the four values of $\beta$ = 0.16, 0.75, 3.51
and 4.89 from the three- and one-parameter fits discussed in the text. These
are marked by vertical dotted lines.}
\label{fig:mod}
\end{figure}

We now turn to the estimate of the $CP$-conserving contribution to $K_L
\rightarrow \pi^0 e^+ e^-$ using the model of Ref.~\cite{doga}. Using
the results of the fit to the shape of the distribution only,
Eqs.~(\ref{avfit})~and~(\ref{3pfit}), we find
\begin{equation}
B_{CPC}(K_L \rightarrow \pi^0 e^+ e^-) = \cases{
4.0 \times 10^{-13} & {\rm vector meson dominance} \cr
2.0 \times 10^{-13} & {\rm three-parameter~fit .} \cr
}
\label{comp}
\end{equation}
Notice that these two numbers are an
order of magnitude smaller than what is obtained using the KTeV data
instead (see Eq.~(11) of Ref.~\cite{gabval}).
We can see from Fig.~\ref{fig:mod} why the NA48 result \cite{na48}
implies a much smaller $B_{CPC}(K_L \rightarrow \pi^0 e^+ e^-)$ than
the KTeV result \cite{ktev} ($\beta = -5$ for the three-parameter fit
or $\beta = 7.5$ for the $a^{\phantom{l}}_V$ fit). These two points
are shown as the two internal dotted lines in Fig.~\ref{fig:mod}. It
is clear from this figure that
the NA48 results correspond to a $K_L \rightarrow \pi^0 \gamma \gamma$
that produces a minimal $CP$-conserving contribution in
$K_L \rightarrow \pi^0 e^+ e^-$, {\it i.e.} it indicates that the two
photons have a negligible D-wave component. The VMD result in Eq.~(\ref{comp})
is consistent with the result reported by NA48. The latter is based on an
analysis of the low $m^{\phantom{l}}_{\gamma \gamma}$ region only and yields
$B_{CPC}(K_L \rightarrow \pi^0 e^+ e^-) =(4.7^{+
  2.2}_{-1.8})\times 10^{-13}$ \cite{na48}. The NA48 result is
obtained from data with $m^{\phantom{l}}_{\gamma\gamma}$ below 110~MeV and is
therefore model independent because in that region the $B$ amplitude
dominates and the correlation with the $A$ amplitude implied by the VMD
ansatz disappears.

If we use the results of the fits to both rate and spectrum,
Eqs.~(\ref{avfitr})~and~(\ref{3pfitr}), we find instead,
\begin{equation}
B_{CPC}(K_L \rightarrow \pi^0 e^+ e^-) = \cases{
(13.8^{+0.9}_{-2.1}) \times 10^{-13} & {\rm vector meson dominance} \cr
(4.6^{+1.7}_{-2.2}) \times 10^{-13} & {\rm three-parameter~fit .} \cr
}
\label{compr}
\end{equation}
These two points are shown as the external dotted lines in Fig.~\ref{fig:mod}.
Not surprisingly, the general three-parameter fit continues to
agree with the model independent NA48 limit as it gives a good fit to
both the rate and spectrum. On the other hand, the fit in terms of
$a^{\phantom{l}}_V$ alone does not reproduce the data very well and we can
dismiss its implication of a larger $B_{CPC}(K_L \rightarrow \pi^0 e^+ e^-)$.

In Fig.~\ref{fig:mod} we see why there are two different solutions for
$\beta$ that result in the same $B_{CPC}(K_L \rightarrow \pi^0 e^+ e^-)$.
This $CP$-conserving component depends
quadratically on the $B(z)$ amplitude of $K_L \rightarrow \pi^0 \gamma
\gamma$, and therefore there are two values of $\beta$ for any given
$B_{CPC}(K_L \rightarrow \pi^0 e^+ e^-)$. They correspond to constructive
and destructive interference between the term with $a^{\phantom{l}}_2$ and $\beta$
in Eq.~(\ref{ampl}).

\section{Conclusion}

We have shown that the NA48 data for the reaction
$K_L \rightarrow \pi^0 \gamma \gamma$ can be accommodated nicely by the
theoretical expression based on chiral perturbation theory at
order $p^6$. With this expression it is possible to describe
simultaneously the total rate and the shape of the spectrum, which is
not possible with chiral perturbation theory at order $p^4$ \cite{earlyex}.
We have also shown that the commonly used VMD ansatz fails in this
case, and that it is impossible to fit both the rate and the shape of
the spectrum if this ansatz is adopted.

We have also shown that it is possible to obtain
a good fit to this mode for different values of the poorly
known parameter $a^{\phantom{l}}_2$ from $K\rightarrow 3\pi$ decays. This indicates
both that $K_L \rightarrow \pi^0 \gamma \gamma$ cannot provide
additional information on the value of $a^{\phantom{l}}_2$, and that not knowing
its precise value does not affect our ability to describe
the features of $K_L \rightarrow \pi^0 \gamma \gamma$.

Although we do not have sufficient information to perform a similar
comparison for the KTeV data, we note that the value of
$a^{\phantom{l}}_V$ reported by KTeV \cite{ktev}, $a^{\phantom{l}}_V =
-0.72 \pm 0.05 \pm 0.06$ predicts a branching ratio $B(K_L \rightarrow
\pi^0 \gamma \gamma) = (1.36 \pm 0.06 \pm 0.07) \times 10^{-6}$ in
conflict with the measured value, $B(K_L \rightarrow \pi^0 \gamma
\gamma) = (1.68 \pm 0.07 \pm 0.08) \times 10^{-6}$.

The new results from NA48
indicate a very small D-wave component for the photon pair
and this leads to a prediction of a negligible $CP$-conserving
background to $K_L \rightarrow \pi^0 e^+ e^-$. We have shown that this
result is not an artifact of the VMD ansatz and that it holds in
the general parametrization.
This result is at odds with the earlier KTeV data and we must
wait for the new KTeV results to see how this discrepancy is resolved.

\section*{Acknowledgments}

\noindent This work was supported in part by DOE under Contract Number
DE-FG02-01ER41155. We thank R. Wanke and M. Martini for helpful
discussions of the NA48 results. We thank James Cochran for bringing
Ref.~\cite{stat} to our attention.

\end{document}